# Delicate competing electronic states in ultrathin manganite films

Zhaoliang Liao[†], Rongying Jin, E. W. Plummer, Jiandi Zhang*

*Department of Physics and Astronomy, Louisiana State University, Baton Rouge, LA 70810, USA*

*E-mail: jiandiz@lsu.edu

[†]Current address: *Materials Science and Technology Division, Oak Ridge National Laboratory, Oak Ridge, Tennessee 37831, USA*

**Abstract**

The coupling between the electrical transport properties of $La_{2/3}Sr_{1/3}MnO_3$ (LSMO) thin films and structural phase transitions of $SrTiO_3$ (STO) substrates at $T_s$ = 105 K has been investigated. We found that the electrical resistivity of LSMO films exhibit a "cusp" at $T_s$, which is greatly amplified by tuning films to the verge of metallic and insulating phases, i.e., to the boundary of two delicate competing electronic states. Our results demonstrate that small amounts of strain can tip the subtle balance of competing interactions and tune the electronic properties in correlated electron materials.

Novel phenomena in complex transition-metal oxides, such as superconductivity, colossal magnetoresistance, and multiferroicity, have attracted great attention for several decades. They are believed to be intimately related to synergistically competing interactions between charge, lattice, spin, and orbital degrees of freedom.[1,2] Thus, the material properties can be extremely sensitive to intrinsic and extrinsic stimulus such as doping, strain, pressure, and field, giving rise to extraordinary behaviors.[3-8] In particular, the rapid development of advanced techniques for oxide heterostructure growth at the atomic level [9,10] provide additional twists of "man-made" dimensions to manipulate the competing interactions. For example, one can obtain interesting quantum phenomena in correlated materials by reducing dimensionality, which has been demonstrated at the interface of insulating compounds $LaAlO_3$ (LAO) and $SrTiO_3$ (STO) where a two-dimensional electron gas is formed.[11] A much higher Néel temperature has been achieved by fabricating cation ordered $(LaMnO_3)_m/(SrMnO_3)_{2m}$ superlattices compared to the compositionally equivalent $La_{1/3}Sr_{2/3}MnO_3$ random alloy.[12] Magnetic anisotropy in perovskite thin films can be controlled on the atomic scale by engineering interface structure [13] and oxygen coordination environment.[14]

As a prototype substrate, STO undergoes a second-order cubic to tetragonal phase transition at $T_s \sim 105$ K.[15] The effect of such an antiferrodistortive phase transition on the ferromagnetism has been observed in $La_{2/3}Sr_{1/3}MnO_3$ (LSMO) thin films grown on STO,[16-18] where magnetization reveals an anomaly across $T_s$. Since the metallicity of LSMO is coupled to its ferromagnetism,[19] the electrical conductivity is expected to change accordingly. However, the impact of the STO structural transition on the electrical transport property of LSMO is rarely reported, in spite of extensive investigation of LSMO thin films on (001) STO. Ref. [20] reports an apparent cusp in the resistivity curve of $La_{0.5}Sr_{0.5}MnO_3$/STO at $T_s$ [20], which is suggested to be related to the strong coupling between the soft phonon mode ($\Gamma_{25}$) [15] of STO and charge carriers (holes) of La$_{0.5}$Sr$_{0.5}$MnO$_3$. On the other hand, the structural transition of STO, which changes the STO lattice constant and thus the strain in the LSMO films, is expected to strongly affect the electrical transport as well.[21] In this article, we demonstrate that the impact of the STO structural transition on the electrical transport property can be significantly amplified in ultrathin LSMO films when approaching the critical condition for the metal-insulator transition (MIT). Our results reveal the characteristic of strongly competing electronic phases in the system, thus offering an opportunity to manipulate physical properties by tuning strain.

High-quality LSMO films were grown on TiO$_2$-layer terminated (001) STO substrates by pulsed laser deposition. Single TiO$_2$ terminated STO substrates were achieved by standard buffer HF acid etching. The details of growth can be found in Ref. 21. The electrical resistivity was measured by a Quantum Design physical properties measurement system (QD-PPMS). With optimized growth conditions under an oxidizing environment, we found a minimum critical thickness ($t_{cr}$) of 6 unit cells (u.c.) for the MIT in LSMO films on STO substrates.[21] Figure 1 presents the temperature ($T$) dependence of the electrical resistivity ($\rho$) of LSMO films on STO substrates grown at an oxygen pressure ($P_O$) of 130 mTorr. As shown in Fig. 1a, the 6 u.c. LSMO film shows a "cusp" in resistivity at $T_s$. It becomes more obvious when taking the derivative of resistivity as can be seen in Fig. 1b. Therefore, the anomaly in resistivity at $T_s$ must be directly related to the substrate phase transition.

Naively, one would expect that the resistivity anomaly occurs in all LSMO/STO films. Figure 1c presents $\rho(T)$ for LSMO/STO films with different thicknesses. It is found that the resistivity cusp solely appears in the film with the critical thickness ($t_{cr}$ = 6 u.c), which is at the boundary between the metallic ($t > t_{cr}$) and insulating ($t < t_{cr}$) phases. The other films do not exhibit an evident cusp at 105 K. The thickness dependence of the $\rho(T)$ anomaly at $T_s$ is remarkable, because in principle the substrate effect should be stronger on the properties of thinner films.[22] Obviously, the observed non-monotonic response with increasing film thickness cannot be simply explained as strain relaxation behavior in our films. The immediate disappearance of the resistivity cusp when away from the critical thickness by only one unit cell indicates that STO structure plays an important role in the film at the verge between the metal and insulator phases.

If the structure-resistivity coupling is enhanced in films that are near the metal-insulator transition (MIT), the resistivity cusp then is expected to be sensitive to film stoichiometry, top interface effects (i.e., the surface of the film), and magnetic field since they can alter the critical thickness $t_{cr}$.[21] To illustrate this, we have investigated the resistivity of LSMO films with different growth oxygen pressures ($P_O$), STO capping layers, and under the application of magnetic field ($H$). To simplify sample labeling, we use ($n$, $P_O$, $H$) to represent an $n$ u.c. thick LSMO film grown at $P_O$ oxygen pressure with its resistivity measured under a magnetic field $H$. The label $nc$ denotes an LSMO film with $n$ u.c. thickness capped with 2 u.c. STO. Figure 2a displays the $T$-dependence of the resistivity of 6.u.c films grown at different oxygen pressures. Note that the resistivity cusp becomes more evident with increasing $P_O$. According to our previous study [21], films grown under reduced oxygen pressure will increase oxygen vacancy density, making the film more insulating. This in turn drives the 6 u.c. film away from the critical thickness that separates the metallic and insulating phases.

Secondly, the resistivity cusp can be tuned by changing the top interface. Figure 2b shows the $T$-dependence of resistivity for films grown under $P_O$ = 130 mTorr and capped by 2 u.c. STO. Compared with the 5 uc film without capping (see the bottom curve of Fig. 1c), the capped 5 u.c. film (5c, 130, 0) reveals a pronounced resistivity cusp, suggesting that the STO capping drives the MIT toward thinner films. The (6c, 130, 0) film, though metallic, has a resistivity one order of magnitude higher than (7, 130, 0) (see the third curve of Fig. 1c), and is therefore also very close to the verge of a MIT from the metallic side. It exhibits a resistivity cusp as well. The (4c, 130, 0) and (7c, 130, 0) films are either too insulating or too metallic to reveal resistivity cusps.

Shown in Fig. 2c is the *T*-dependence of resistivity for the uncapped films under various magnetic fields. Note that applying a 14 T magnetic field makes the 5 u.c. LSMO film (5, 130, 14) less insulating than the zero field case and leads to a resistivity cusp at $T_s$ (see the bottom curve of Fig. 1c where no cusp appears). For the same reason, the resistivity cusp in the 6 u.c. LSMO film is also enhanced (Fig. 2c) under the application of magnetic field. Particularly, there is a change in cusp shape when the field increases from 7 T to 14 T, even though the magnetization of the film is already saturated under 7 T field. This suggests that the field effect should be more intrinsically related to magneto-elastic and/or double-exchange coupling rather than a magnetic domain structure effect. These facts indicate that the magnetic field drives both the 5 u.c. and 6 u.c. films closer to the verge of a MIT, thus exhibiting more obvious resistivity cusps at $T_s$.

The above results clearly demonstrate that film thickness, oxygen concentration, top interface modulation (capping), and magnetic field can significantly modify the film MIT boundary, reflected by the resistivity cusp. To further quantify the effect induced by the STO structural transition, we analyze the slope of resistivity just below ($k_1$) and above ($k_2$) $T_s$. To describe the sharpness of a cusp, we introduce the angle $\Theta \equiv \tan^{-1}(k_2/k_2)-\tan^{-1}(k_1/k_2) = 45°-\tan^{-1}(k_1/k_2)$ shown in Fig. 3a. When $\Theta = 0$ ($k_1 = k_2$), there is no cusp, and larger $\Theta$ means a stronger cusp. We also use a slope $\beta$ in the Arrhenius plot of resistivity at temperatures higher than $T_s$ for the insulating films shown in Fig. 3b. Linear behavior is observed in the ln$\rho$-1000/$T$ plot at $T > T_s$. If thermal activation dominates the electrical transport, the slope from the Arrhenius plot indicates the thermal activation energy ($\Delta E = k_B \times \beta$, where $k_B$ is the Boltzmann constant). Therefore, a larger $\beta$ implies a higher energy gap, i.e., more insulating. Fig. 3c shows the Arrhenius plot of resistivity for various insulating LSMO films. The nice linear behavior allows us to accurately determine the $β$ value, e.g., $β$ ((6, 180,0)) < $β$ ((6, 130,0)) < $β$ ((4, 180,0)).

The effects of film thickness, oxygen pressure, STO capping, and external magnetic field on the resistivity cusp can be summarized by plotting $Θ$ versus $β$ as shown in Fig. 3d. Note that $Θ$ increases drastically with decreasing $β$. The highest $Θ$ is expected to occur as $β \to 0$. Since $β$ represents the energy gap of an insulating film, $β \to 0$ indicates that the film approaches the verge of MIT. The greatest effect of the STO structural transition on the LSMO transport properties occurs when an LSMO film is located at the verge of a MIT. With increasing $β$, the effect of the STO phase transition becomes weaker and the cusp eventually disappears ($Θ = 0$) when $ΔE$ is too high (> 22 meV). It is concluded that all different tuning parameters (thickness, oxygen stoichiometry, STO capping, and magnetic field) affect the resistivity cusp by varying the thermal activation energy.

As discussed above, the change of the lattice structure of STO at $T_s$ [15] is the origin of the observed resistivity cusp in LSMO ultrathin films. Using a 6 u.c. film grown at $P_O$ = 130 mTorr as an example, Fig. 4a shows that there is a very narrow transition region (from 105 K to 97 K) before and after which the film exhibits typical Arrhenius thermally activated transport behavior. The STO structural transition reduces the thermal activation energy $ΔE$ from 5.72 meV at higher temperature ranges (> 105 K) to 4.13 meV at lower temperatures (< 97 K). Therefore, the lattice effect, which drives the film to be more conductive at lower temperatures, must be considered. The change of the lattice constants of STO during the transition should play a pivotal role in the observed cusp in our LSMO films.

The thermal expansion coefficients of STO (~ $1.11 \times 10^{-5}$ $K^{-1}$) [23] and LSMO (~ $1.16 \times 10^{-5}$ $K^{-1}$) [24] are very close, hence the lattice mismatch should not change too much when changing temperature. However, as shown in Fig. 4b, the in-plane lattice constant decreases more quickly in the tetragonal phase than cubic phase with decreasing temperature.[25] Therefore, STO in the tetragonal phase has a smaller lattice mismatch with LSMO than that in cubic, in favor of the metallic state.[21] This reduced strain explains the sudden change of resistivity below $T_s$. The external magnetic field and the STO capping layer also enhance the conductivity of LSMO films.[21] In contrast, the oxygen deficiency, reduced thickness, and strain, either tensile or compressive, can drive the film to be more insulating. These effects are competing, and the ground state is determined by the delicate balance between several electronic states with close energies. The film with zero thermal activation energy is exactly at such a delicate balance. Any external fluctuation will tip this subtle balance, resulting in a change of electronic properties. An enhanced effect of STO structural transitions on the electrical transport properties have also been reported in other manganite thin films when those films approach insulating phases.[26] A similar situation occurs in zero bandgap materials which also show giant responses to external stimuli.[27,28] These facts therefore suggest a more universal phenomenon due to delicate competing electronic states.

In summary, we demonstrate that the strain variation due to the STO substrate structural transition unambiguously generates considerable effects on the properties of LSMO ultrathin films. The resistivity anomaly, identified as a "cusp", is found to be greatly amplified when the LSMO film is at the verge of a MIT which can be driven by varying film thickness, oxygen stoichiometry, adding a capping layer, and/or applying an external magnetic field. Our results show that small strain can tip the subtle balance of competing interactions and strongly manipulate the physical properties of such a correlated electron material.

**Acknowledgements**

This work was supported by U.S. DOE under Grant No. DOE DE-SC0002136.

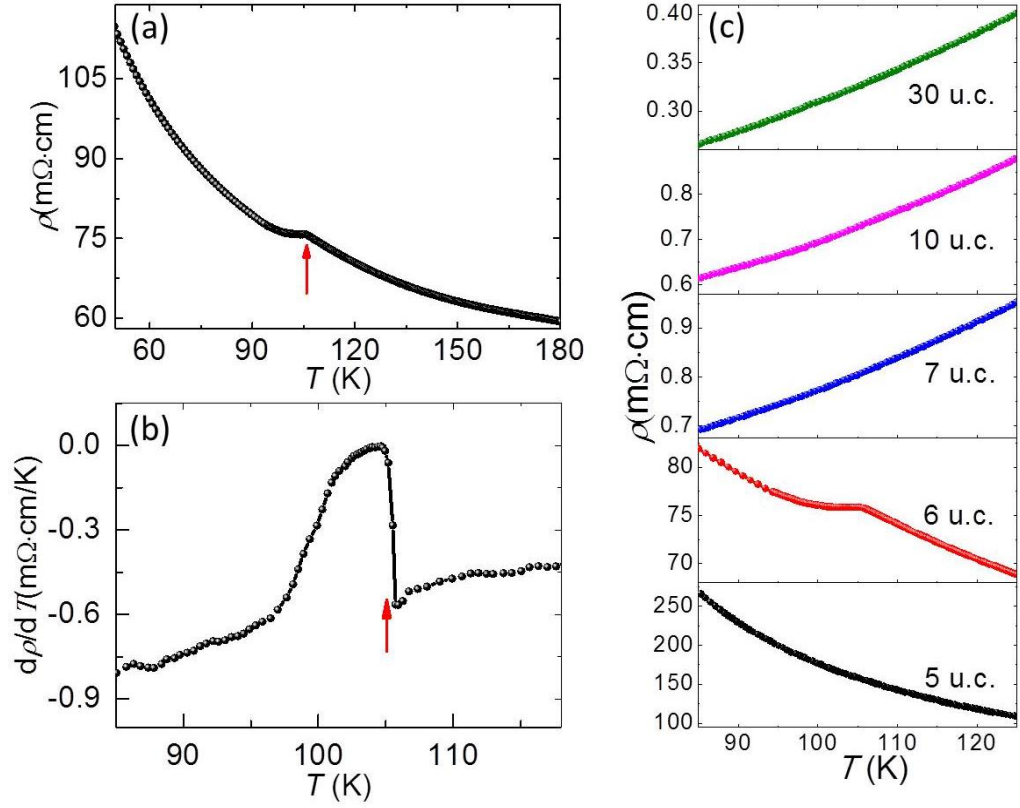

**Figure 1**. (a-b) Temperature (*T*) dependence of the electrical resistivity (a) of the 6 u.c. LSMO film grown at 130 mTorr and its derivative (b). The arrows indicate the transition temperature $T_s$. (c) Temperature dependence of the electrical resistivity in the vicinity of $T_s$ = 105 K for films with different thickness grown at 130 mTorr.

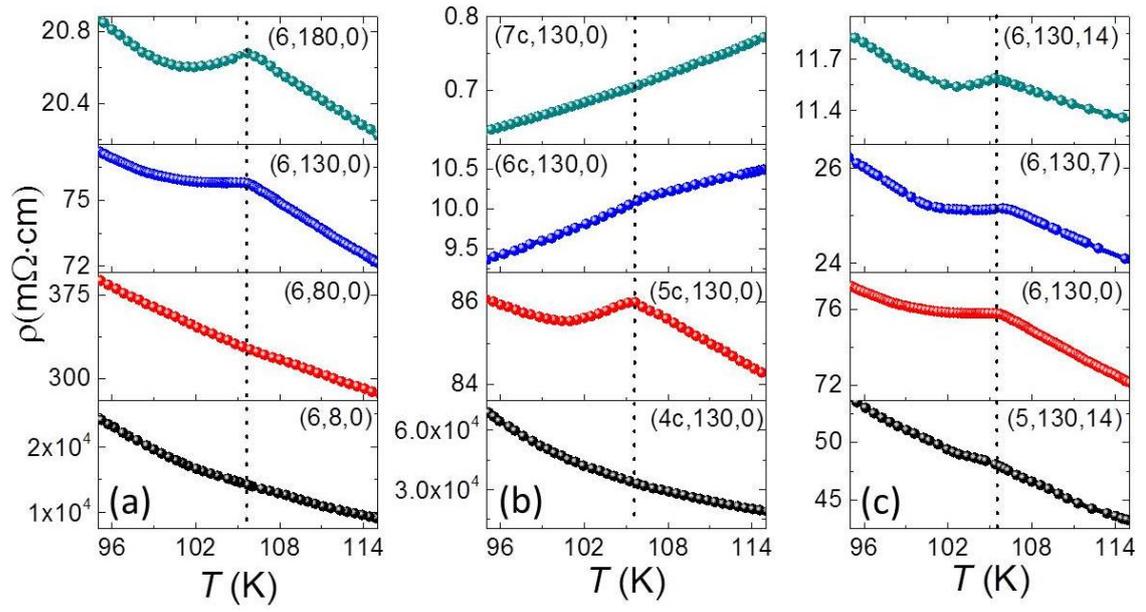

**Figure 2**. *T*-dependent resistivity of (a) 6 u.c. LSMO films grown at different oxygen pressure; (b) different thickness LSMO films capped by 2 u.c. STO grown at 130 mTorr; (c) 5 u.c. and 6 u.c. LSMO films under different magnetic field. The magnetic field was normal to the film surfaces.

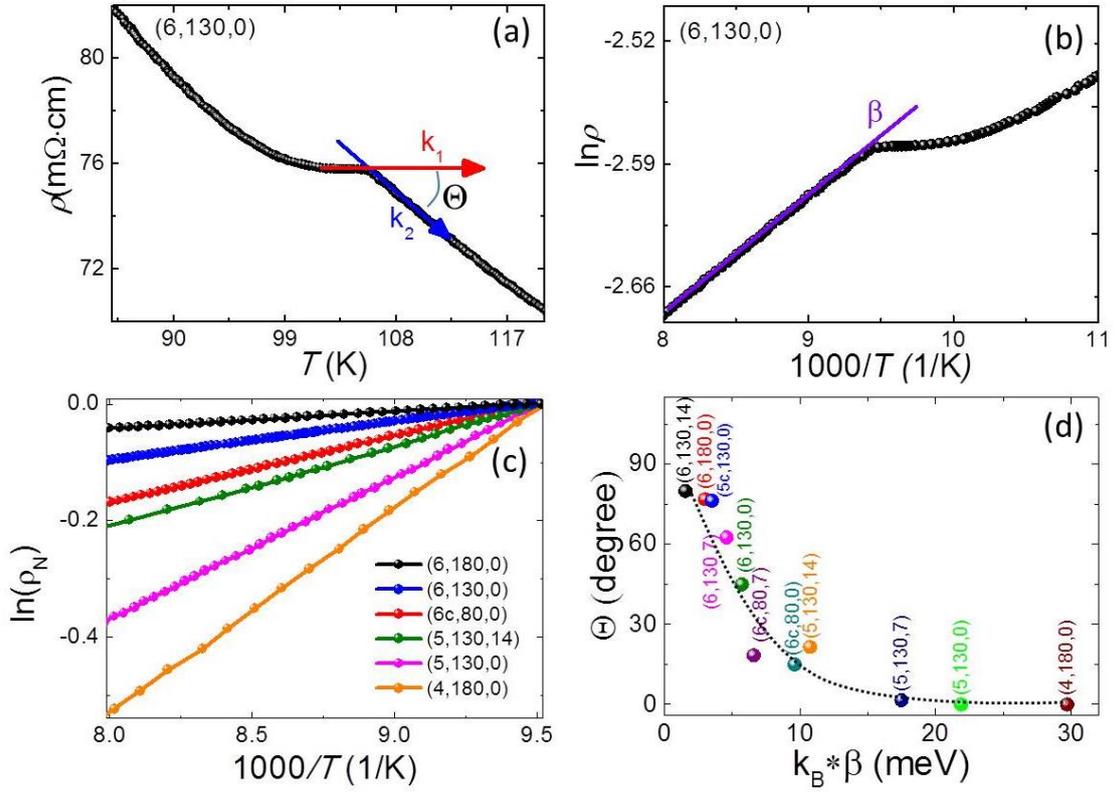

**Figure 3.** (a) Definition of the turning angle $\Theta$. $k_1$ and $k_2$ are the first derivative of resistivity over temperature from below ($\left.\frac{d\rho}{dT}\right|_{0-}$) and above ($\left.\frac{d\rho}{dT}\right|_{0+}$) $T_s$, respectively. $\Theta$ is the angle between $k_2$-normalized $k_1$ and $k_2$ slope lines, i.e., $\Theta \equiv 45°-\tan^{-1}(k_1/k_2)$. (b) The definition of $\beta$, which is the slope of the curve in the Arrhenius plot $\ln\rho$ vs. $1/T$. The sample (6, 130, 0) is used for a demonstration for (a) and (b). (c) Arrhenius plot for a series of insulating LSMO samples. (d) The $\Theta$ as a function of $k_B\beta$, where $k_B$ is Boltzmann constant.

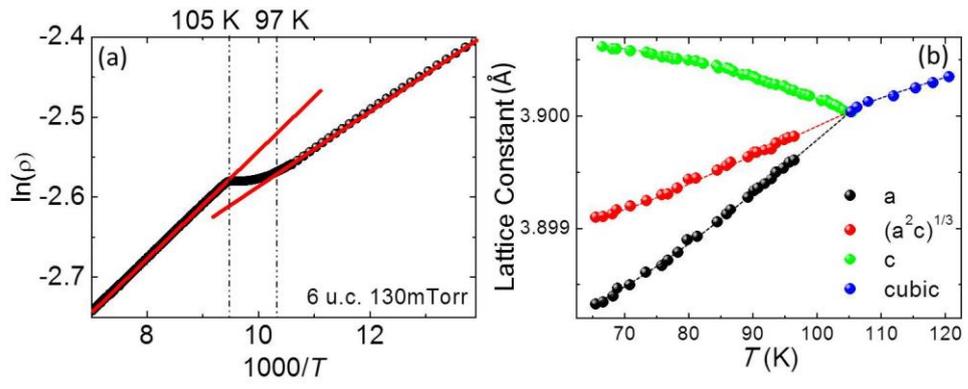

**Figure 4**. (a) ln($\rho$) vs. 1000/$T$ near $T_s$ = 105 K for 6 u.c LSMO film grown at $P_O$ = 130 mTorr. (b) Lattice constants of STO versus temperature [after Ref. 25].